\documentclass[a4paper,11pt]{article}
\pdfoutput=1

\usepackage{jheppub} 

\usepackage[T1]{fontenc} 
\usepackage{graphicx}
\usepackage{epsfig}
\usepackage{bm}
\usepackage{verbatim}
\usepackage[utf8]{inputenc}
\usepackage{booktabs}
\usepackage{multirow}
\usepackage{subfigure}
\usepackage{xcolor}
\usepackage{float}
\usepackage{blindtext}
\usepackage{mathrsfs}

\newcommand{\paren}[1]{\left(#1\right)}
\newcommand{\evalue}[1]{\left.#1\right\vert}
 
\newcommand{\blue}[1]{\textcolor{blue}{#1}}
\newcommand{\magenta}[1]{\textcolor{magenta}{#1}}

\title{\boldmath \magenta{Improvement of W Boson Hadronic Decay Width up to $\mathcal{O}\left(\alpha_s^4 \right)$-order}}

\author[a,1]{Daniel Salinas-Arizmendi,\note{Corresponding author.}}
\author[a]{Claudio Dib,}
\author[a]{and Iván Schmidt}

\affiliation[a]{Departamento de Física, Universidad T\'{e}cnica Federico Santa Mar\'{\i}a, y Centro Científico-Tecnológico de Valparaíso,
Casilla 110-V, Valpara\'{\i}so, Chile}

\emailAdd{daniel.salinas@usm.cl}
\emailAdd{claudio.dib@usm.cl}

\abstract{The principle of maximum conformality (PMC) is used to remove uncertainties in the renormalization scale and scheme, thus eliminating unnecessary systematic errors for high-precision perturbative Quantum Chromodynamics (pQCD) predictions. In this work, we use PMC method to improve the determination of the decay width of the $W$ boson in hadrons by working on the corrections coming from pQCD theory at the four-loop level. In conventional scale setting, we find that the initial scaling dependence of the renormalization is small only at high correction orders, whereas in single-scale PMC method we obtain a scaling-independent initial result.
Finally, the wide hadronic decay of $W$ is used for an indirect determination of the charm-strange mixing parameter.
}


\begin{document} 
\maketitle

\flushbottom

\section{\label{intro}Introduction}
The study of the decay width of the $W$ boson plays an important role within the Standard Model and has been measured by several collaborations, such as D$0$ \cite{D0:1995gzy} ATLAS \cite{ATLAS:2016nqi}, LHCb \cite {LHCb:2016zpq} and CMS \cite{CMS:2022mhs}, where the average leptonic decay branching fraction per flavor is estimated to be $(10.89 \pm 0.08)\%$ and the inclusive hadronic decay branching fraction is $(67.32 \pm 0.23 )\%$.

Improving the predictions of the hadronic decay of the $W$ boson directly affects the information of the fundamental free parameters of the theory. This decay can be used as an excellent estimator of the strong coupling constant, $\alpha_s(Q)$, which determines the scale of the strong interactions theoretically described by Quantum Chromodynamics (QCD); this estimation  has implications for reducing the theoretical uncertainties in the calculations of all high-precision QCD perturbative processes.
In addition, it can be used indirectly to determine the sum of the first two rows of the Cabibbo–Kobayashi–Maskawa (CKM) matrix and thus help
verify the unitarity of this matrix for all quarks lighter than the top quark. Theoretical predictions of the hadronic decay width of the $W$ boson have improved in the last decades due to the incorporation of electroweak (EW) and mix (EW-QCD) corrections, although QCD-type corrections still introduce uncertainties under the premise of renormalization group invariance \cite{Wu:2014iba}.

The aim of this article is to improve the prediction of the decay width of the $W$ boson by an optimal treatment of the perturbative quantum chromodynamics (pQCD) corrections. Conventionally, the renormalization scale of the process, $\mu_r$,  is chosen to be equal to the momentum transfer; in principle a physical result should be independent of $\mu_r$, but due to the truncation of the perturbative series, $\mu_r$ appears in the calculated expressions. This chosen value of $\mu_r$ thus  
poses a problem of ambiguity in the choice of scale and renormalization scheme. An original and heuristic procedure for eliminating ambiguity is the Principle of Maximum Conformality (PMC) method \cite{Brodsky:2011ig}, which has improved the estimates of several high-energy processes. 
The general PMC determines the renormalization scale by absorbing all non-conformal terms governing the running coupling behavior, and thus the resulting pQCD series is conformal.

A first approach of the PMC method is called Brodsky-Lepage-Mackenzie/Principle of Maximum Conformality (BLM/PMC) \cite{Brodsky:2011ta,Brodsky:2012rj}, where all terms containing $N_f$ (number of active quarks) 
in the calculation of a given observable can be absorbed in the definition of the running coupling. The BLM/PMC scales can be determined in a general scheme-independent way and can be different for each order in the perturbative series. Loop by loop we can change the renormalization scales to build a perturbative series with conformal coefficients.

The second approach is to introduce the $\mathcal{R}_\delta$-scheme \cite{mojaza2013systematic,brodsky2014systematic,singlescale1}, which is a generalization of the conventional scheme used in dimensional regularization \citep{bollini1972dimensional,veltman1972regularization},
 in which a constant, $\delta$, is subtracted in addition to the standard $\ln 4\pi - \gamma_E$ subtraction of the $\overline{\text{MS}}$ scheme. The $\delta$ subtraction defines an infinite set of renormalization schemes in which the physical results are independent of $\delta$, i.e. scheme-independent. 

Both approaches come from a generalization of the BLM method \cite{Brodsky:1982gc}, so  the properties of BLM are maintained. Since the PMC predictions do not depend on the choice of the renormalization scheme, the scale of PMC satisfies the principles of RGE invariance \cite{Brodsky:2012ms}.

The rest of the paper is organized as follows: In section \ref{sec:W-decay}, we present the treatment of QCD corrections in inclusive hadronic decay using the PMC method. In section \ref{sec:numerical}, we present our numerical results for $W$ decays in hadrons and compare Conventional Scale Setting and PMC Method. In section \ref{sec:blm}, a brief review of other methods to optimize the choice of renormalization scale and a short explanation of the BLM/PMC method applied to the physical process of this work is presented. Section \ref{sec:ckm} contains the indirect determination of the charm-strange mixing parameter. The section \ref{summary} is reserved for the summary.

\section{\label{sec:W-decay}The $W$ boson hadronic decay width using the PMC}

The hadronic decay of the $W$ boson can be written as \cite{CMS:2022mhs}:
\begin{equation} \label{Wdecay}
\Gamma_W^{\text{(had)}}=\Gamma_0 +\Gamma_{\text{QCD}}+\Gamma_{\text{EW}}+\Gamma_{\text{mix}},
\end{equation}
%
where $\Gamma_0$ denotes the tree level contribution. In the massless quark limit, it is:
\begin{equation} \label{Wborn}
\Gamma_0=\frac{\sqrt{2} G_F N_{c}}{12 \pi} M_{W}^3 \sum_{i=u,c}\ \sum_{j=d,s,b}\left|V_{i j}\right|^2,
\end{equation}
which depends on the color number $N_c$ ($=3$ for quarks), the Fermi constant $G_F$, the mass of the $W$ boson, and the sum of the square of the CKM matrix elements, $V_{ij}$, excluding the kinematically prohibited terms (mixed terms with the top). 
The term $\Gamma_{\text{QCD}}$ includes the pure QCD corrections, which will be the focus of this work. On the other hand, $\Gamma_{\text{EW}}$ are the electroweak corrections, found  in \cite{Bardin:1986fi,Denner:1990tx,Denner:1991kt,Kniehl:2000rb}, and $\Gamma_{\text{mix}}$ corresponds to EW-QCD mixed corrections \cite{Kara:2013dua,dEnterria:2016rbf,dEnterria:2020cpv}, i.e.\ the corrections proportional to the square of the quark to boson mass ratio, which are strongly suppressed compared to the other corrections.

The QCD corrections of the $W$ boson decay to hadrons can be expressed as

\begin{equation}\label{eq:rNS}
\Gamma_{\text{QCD}}=\Gamma_0 \sum_{n = 1}^\infty \sum_{m=0}^{n-1} c_{n,m}\,  {N_f}^m \ a_s^n,
\end{equation}
%
where $a_{s}=\alpha_{s}/(4\pi )$, the coefficients $c_{n,m}$ arise from loop corrections and are known up to fourth order in $a_s$ \cite{Baikov:2008jh},  the subindex $n$ is the order of the perturbation (power of $a_s$), ,$m$ the number of internal fermionic loops, and $N_f$ is the number of active flavors (quarks lighter than the transferred momentum).

In order to adequately deal with ultraviolet (UV) divergences in $\Gamma_W^{(\text{had})}$ arising from quantum corrections, 
we apply dimensional regularization with $D=4-2\epsilon$ \cite{bollini1972dimensional,veltman1972regularization} and introduce a kinematic scale $\mu$. The divergences appear as poles $1/\epsilon$, which can be absorbed in the renormalization of the coupling
$a_s$: the bare strong coupling, $a_s^{(0)} \equiv {g_s^{(0)}}^2/16\pi^2$ is given in terms of the renormalized gauge coupling $a_s^{R}$ as:
\begin{equation}
a_s^{(0)} = \mu^{2\epsilon} Z_a a_s^R,
\end{equation}
where the scale $\mu$ is introduced to preserve $a_s^R$ dimensionless, and $Z_a$ is the renormalization factor 
obtained from the generating function of the bare Green function; in our case it can be written as
\begin{equation}
\begin{aligned}
 Z_a-1= \delta_a = & \sum_{i=1}^{4} \sum_{j=-i}^{0} a_s^{i} \frac{c_i^{[j]}}{\epsilon ^{-j}}\\
 = & \ -\frac{\beta_0}{\epsilon} a_s+\left(\frac{\beta_0^2}{\epsilon^2}-\frac{\beta_1}{2 \epsilon}\right) a_s^2-\left(\frac{\beta_0^3}{\epsilon^3}-\frac{7 \beta_0 \beta_1}{6 \epsilon^2}+\frac{\beta_2}{3 \epsilon}\right) a_s^3 \\
& +\left(\frac{\beta_0^4}{\epsilon^4}-\frac{23 \beta_1 \beta_0^2}{12 \epsilon^3}+\frac{5 \beta_2 \beta_0}{6 \epsilon^2}+\frac{3 \beta_1^2}{8 \epsilon^2}-\frac{\beta_3}{4 \epsilon}\right) a_s^4+ \mathcal{O}\left(a_s^5\right),
\end{aligned}
\end{equation}
where the coefficients $c_i^{[j]}$ are given explicitly up to order $i=4$ in the expression, in terms of the $\beta_i$, which in turn are the coefficients of the expansion of the $\beta$ function, defined in \cite{Tarasov:1980au}: 
\begin{eqnarray}
 \beta_0 &= &11-\frac{2}{3}N_f, \\
 \beta_1 &= &102 - \frac{38}{3} N_f, \\
 \beta_2^{\overline{\text{MS}}} &= & \frac{2857}{2} - \frac{5033}{18}N_f + \frac{325}{54}N_f^2, \\
 \beta_3 ^{\overline{\text{MS}}} &= & \left(\frac{149753}{6}+3564 \zeta_3\right)-\left(\frac{1078361}{162}+\frac{6508}{27} \zeta_3\right) N_f \\ 
\nonumber & & +\left(\frac{50065}{162}+\frac{6472}{81} \zeta_3\right) N_f^2+\frac{1093}{729} N_f^3.
\end{eqnarray}

The bare coupling  is independent of the arbitrary scale $\mu$, and therefore  $a_s^R$ (henceforth simply referred to as  $a_s$) must obey the differential equation of the renormalization group (RGE), given by:
\begin{equation}\label{RGEs}
\frac{da_s}{d\ln \mu^2} =\beta\paren{a_s} =-\sum_{i \geq 0} \beta_i\, a_s^{i+2}.
\end{equation}

By integrating perturbatively Eq.~\eqref{RGEs}, we can introduce a scale displacement $\mu_0\to \mu^\ast$  for the strong coupling up to five-loop level:
\begin{equation}
\begin{aligned}
a_s\left(\mu_*\right)  = &\ a_s(\mu_0)-\beta_0 \ln \left(\frac{\mu_*^2}{\mu_0^2}\right) a_s^2(\mu_0)+\left[\beta_0^2 \ln ^2\left(\frac{\mu_*^2}{\mu_0^2}\right)-\beta_1 \ln \left(\frac{\mu_*^2}{\mu_0^2}\right)\right] a_s^3(\mu_0) \\
 & +\left[-\beta_0^3 \ln ^3\left(\frac{\mu_*^2}{\mu_0^2}\right)+\frac{5}{2} \beta_0 \beta_1 \ln ^2\left(\frac{\mu_*^2}{\mu_0^2}\right)-\beta_2 \ln \left(\frac{\mu_*^2}{\mu_0^2}\right)\right] a_s^4(\mu_0)\\
 & + \left[ \beta_3 \ln \left(\frac{\mu_*^2}{\mu_0^2}\right) +\frac{3}{2}\beta_1^2 \ln^2 \left(\frac{\mu_*^2}{\mu_0^2}\right)+3\beta_0 \beta_2 \ln^2 \left(\frac{\mu_*^2}{\mu_0^2}\right) - \frac{13}{3} \beta_0^2 \beta_1 \ln^3 \left(\frac{\mu_*^2}{\mu_0^2}\right)\right.\\
 & + \beta_0^4 \ln^4 \left(\frac{\mu_*^2}{\mu_0^2}\right)\bigg]a_s^5(\mu_0) + \mathcal{O}\left(a_s^6\right). 
\end{aligned}
\end{equation}   
\begin{figure}
\centering
\includegraphics[scale=0.65]{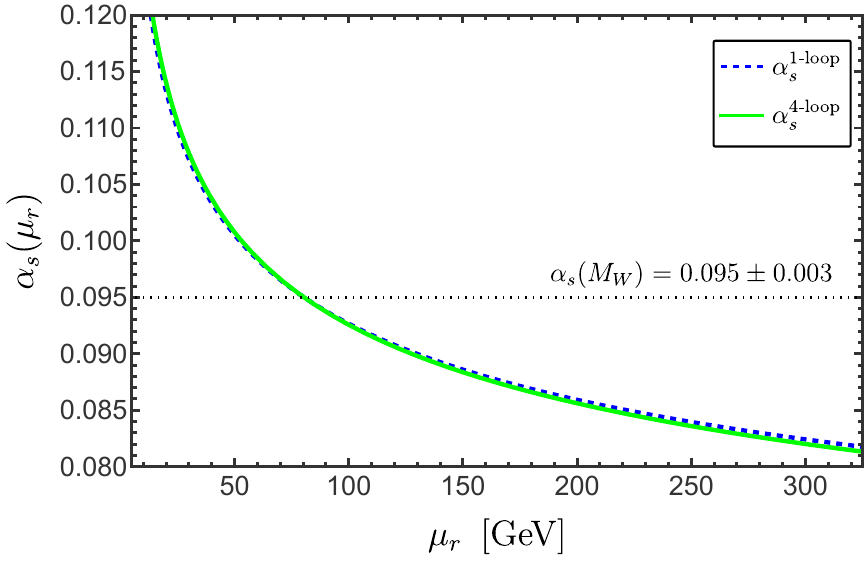}
\caption{The running strong coupling constant $\alpha_s(\mu_r)$ as a function of the renormalization scale $\mu_r$. The plot shows the evolution of $\alpha_s$ at the $1$-loop (blue dashed curve) and $4$-loop (green solid curve) approximations. 
The horizontal dotted line represents the reference value of $\alpha_s$ at the $W$ boson mass \cite{Workman:2022ynf}. 
}
    \label{fig:strongcoupling}
\end{figure}

Figure \ref{fig:strongcoupling} illustrates the running of the strong coupling constant $\alpha_s(\mu_r)$ as a function of the renormalization scale $\mu_r$ for both the 1-loop (dashed blue curve) and 4-loop (solid green curve) approximations. 
The evolution of $\alpha_s(\mu_r)$ demonstrates the asymptotic freedom property of QCD, wherein the coupling decreases with increasing energy scale, reflecting the reduced interaction strength between quarks at higher energies. The horizontal dotted line represents the reference value of the strong coupling at the $W$ boson mass, $\alpha_s(M_W) = 0.095 \pm 0.003$, illustrating the consistency of the 4-loop prediction with the established input.

As shown in a previous work \cite{Salinas-Arizmendi:2022pzj}, using the scale displacement relation, the QCD correction $\Gamma_{\text{QCD}}$ in Eq.\eqref{eq:rNS} can be expressed as
\begin{equation}\label{eq:deltaqcd}
\Gamma_{\text{QCD}}\paren{Q}= \Gamma_0 
\sum_{n = 1}^\infty \sum_{m=0}^{n-1}
%
\Bigg( r_{n,0}\ +\ n(-1)^{m}\ r_{n+m,m}(Q/\mu) \ \beta(a_s) \  B_{n}^{(m-1)}\, a_s^{-1}(\mu) \Bigg) a_s^{n}(\mu),
\end{equation}
where the first three factors $B_n\paren{a_s}$ are:
\begin{equation}
\begin{aligned}
& B_n^{(0)}=1, \quad B_n^{(1)}=\frac{1}{2}\left[\frac{\partial \beta}{\partial a_s}+(n-1) \frac{\beta}{a_s}\right], \\
& B_n^{(2)}=\frac{1}{3!}\left[\beta \frac{\partial^2 \beta}{(\partial a_s)^2}+\left(\frac{\partial \beta}{\partial a_s}\right)^2+3(n-1) \frac{\beta}{a_s} \frac{\partial \beta}{\partial a_s}+(n-1)(n-2) \frac{\beta^2}{a_s^2}\right].
\end{aligned}
\end{equation}
Here $r_{n,0}$ are adimensional numbers, called conformal coefficients, 
and $r_{n+m,m}(Q/\mu)$ are called non-conformal coefficients, which are functions
of $Q/\mu$. 
These coefficients are uniquely related to the  $c_{n,m}$ of Eq.~\eqref{eq:rNS}, as shown in Ref.~\cite{brodsky2014systematic}, and  are  of the form:
\begin{equation}
r_{n+m,m}(Q/\mu) =\sum_{k=0}^{m} C_k^m\  \hat{r}_{n+m-k,m-k}\ \ln ^{k}\left( \frac{\mu^{2}}{Q^{2}}%
\right),
\end{equation}
with $C_k^m=m!/\left[ k!\paren{m-k}!\right]$. 
The numerical coefficients $\hat{r}_{n+m-k,m-k}$ are shown in \hbox{Appendix \ref{ape1}}.
Notice in particular that $r_{n,0}=\hat{r}_{n,0}$ is purely numeric, independent of $Q/\mu$.

Using the PMC method \cite{singlescale1} in the single-scale approach (PMCs) we obtain a result of maximum conformality pQCD series for $\Gamma_{\text{QCD}}$:
\begin{eqnarray} \label{rnsPMC}
\left. \Gamma_{\text{QCD}}\right\vert _{\text{PMCs}} &=& \Gamma_0 \bigg( \hat{r}_{1,0}\ a_{s}\left(
Q_{s}\right) +\hat{r}_{2,0}\ a_{s}^2\left( Q_{s}\right) +\hat{r}_{3,0}\ a_{s}^{3}\left( Q_{s}\right) \\
&&  +\hat{r}_{4,0}\ a_{s}^{4}\left( Q_{s}\right)+%
\mathcal{O}\left( a_{s}^{5}\right)  \bigg) .  \nonumber
\end{eqnarray}%
This expression is of \emph{maximum conformality} in the sense that it depends on the $\hat r_{n,0}$ only, and not on the $r_{n+m,m}(Q/\mu)$, provided the right $Q_s$ value is used.  $Q_{s}$\ is the PMCs scale, 
which is determined by the
following perturbative series:
\begin{equation}\label{single-scale}
\ln \frac{Q_{s}^{2}}{M_{W}^{2}}=T_{0}+T_{1}a_{s}\left( M_{W}\right)
+T_{2}a_{s}^2\left( M_{W}\right) +\mathcal{O}\left( a_{s}^{3}\right),
\end{equation}%
with
\begin{eqnarray}
T_{0} &=&\frac{\hat{r}_{2,1}}{\hat{r}_{1,0}}, \\
T_{1} &=&\frac{\left( \hat{r}_{2,1}^{2}-\hat{r}_{1,0}\hat{r}_{3,2}\right)
\beta _{0}}{\hat{r}_{1,0}^{2}}+\frac{2\left( \hat{r}_{2,0}\hat{r}_{2,1}-\hat{%
r}_{1,0}\hat{r}_{3,1}\right) }{\hat{r}_{1,0}^{2}},\\
T_{2} &=&\frac{3\left( \hat{r}_{2,1}^{2}-\hat{r}_{1,0}\hat{r}_{3,2}\right) 
\beta _{1}}{2\hat{r}_{1,0}^{2}} +\frac{\left( 2\hat{r}_{1,0}\hat{r}_{3,2}\hat{r}_{2,1}-\hat{r}_{2,1}^{3}-%
\hat{r}_{4,3}\hat{r}_{1,0}\right) \beta _{0}^{2}}{\hat{r}_{1,0}^{3}}\\
 \nonumber && +\frac{3\left( \hat{r}_{1,0}\hat{r}_{2,1}\hat{r}_{3,0}-\hat{r}_{1,0}^{2}%
\hat{r}_{4,1}\right) +4\left( \hat{r}_{1,0}\hat{r}_{3,1}\hat{r}_{2,0}-\hat{r}%
_{2,1}\hat{r}_{2,0}\right) }{\hat{r}_{1,0}^{3}}  \nonumber \\
\nonumber &&+\frac{\left( 4\hat{r}_{1,0}\hat{r}_{2,1}\hat{r}_{3,1}-3\hat{r}_{1,0}^{2}%
\hat{r}_{4,2}+2\hat{r}_{1,0}\hat{r}_{3,2}\hat{r}_{2,0}-3\hat{r}_{2,1}^{2}%
\hat{r}_{2,0}\right) \beta _{0}}{\hat{r}_{1,0}^{3}}.
\end{eqnarray}

The single scale 
$Q_s$ chosen by the PMC method is almost independent of the choice of the initial renormalization scale $\mu_r$, which is the merit of this method.
As a result, the QCD correction for the \( W \) boson in the PMCs method, as obtained in \eqref{rnsPMC}, is similar to the expansion in \eqref{eq:deltaqcd}, but contains only conformal terms, with $\beta = 0$, and is therefore less dependent on the initial renormalization scale, which reduces the ambiguity in the choice of the renormalization scheme. In the following section, the obtained predictions will be studied numerically.


\section{\label{sec:numerical}Numerical results}

For our numerical analysis, we use the following input parameters: the $W$ boson mass $M_W=80.377\pm 0.012$ GeV,  the strong coupling $\alpha_s\paren{M_Z}=0.1179 \pm 0.0009$, and the $Z$ boson mass $M_Z=91.1876 \pm 0.0021\ \text{GeV}$ \cite{Workman:2022ynf}.

\begin{figure}[]
\centering
\subfigure[]{\includegraphics[scale=0.51]{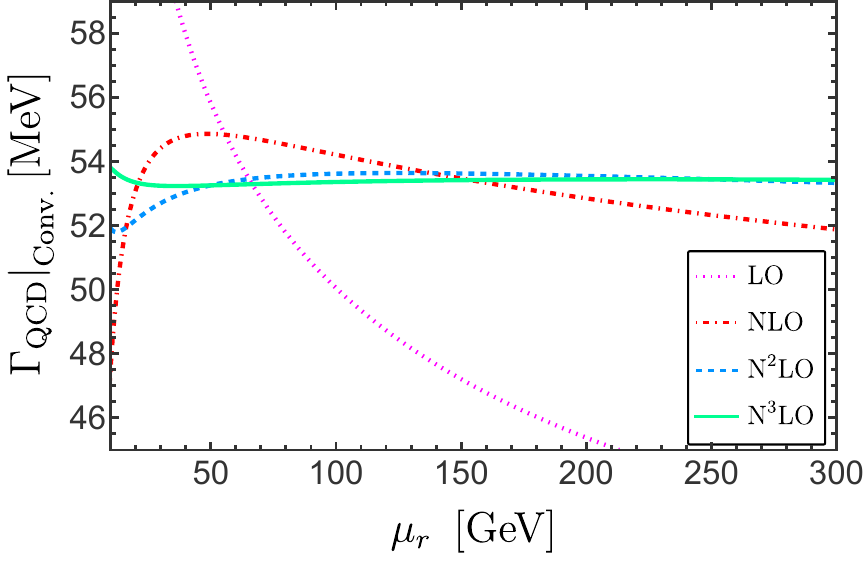}}
\subfigure[]{\includegraphics[scale=0.51]{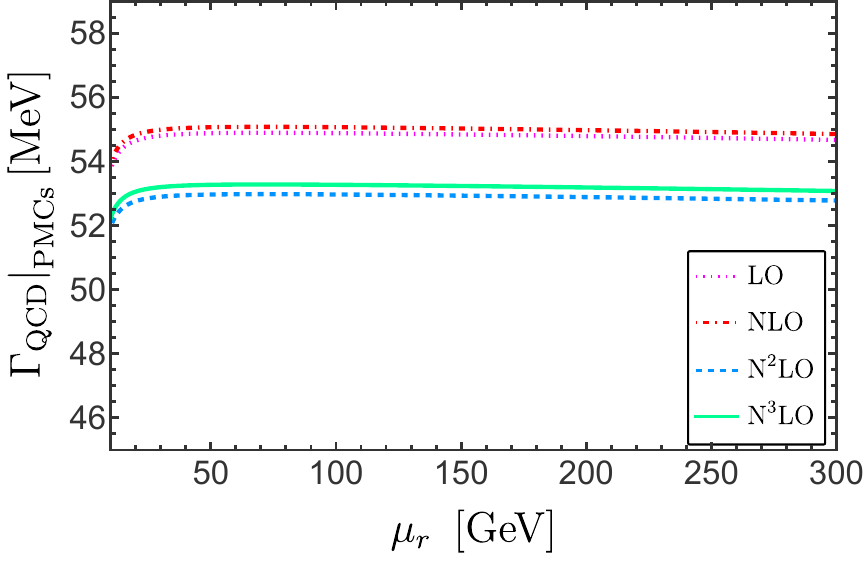}}
\caption{(a) The up to four-loop contributions to  $\Gamma_{\text{QCD}}$ in the conventional setting scale and (b) in the PMC setting scale. We chose the number of quark flavors  $N_f=5$. The input parameters can be found in Section III.}
\label{fig1}
\end{figure}

In the conventional setting of the renormalization scale, the convergence of the perturbative series of  $\Gamma_{\text{QCD}}$ is better if we choose the kinematic scale equal to the renormalization scale, $\mu_r^{init}=\mu_r$, in our case typically equal to $M_W$, so as to avoid large logarithms of the scale ratios.  Although this choice succeeds in suppressing large logarithms and introduces a scale dependence, typically this conventional choice leads to an ambiguity.
Figure \ref{fig1}\blue{(a)} shows the scale dependence of $\Gamma_{\text{QCD}}$ on the conventional scale 
$\mu_r$, with $N_f = 5$, at different loop orders. At leading order (LO) we see a strong dependence on the initial scale (the dotted curve), and only by including further loop corrections the dependence decreases. In this figure the LO in QCD is the magenta dotted line, the next-to-leading order (NLO) is the red dash-dotted line,  the next-to-next-to-leading order ($\text{N}^2$LO) is the blue thick dotted line, and the next-to-next-to-next-to-leading order ($\text{N}^3$LO) is the green solid line. 

On the other hand, the PMC method systematically devises  a way to reduce as much as possible the ambiguity of the choice of $\mu_r$ by optimally determining the effective scale $Q_s$ that suppresses the dependence of the nonconformal parts in $\Gamma_{\text{QCD}}$ at each order of the perturbation theory. The application of the PMC method to $\Gamma_{\text{QCD}}$ is shown in Figure \ref{fig1}\blue{(b)}, where we see that from the LO (magenta dotted curve) there is little scale dependence for regions $\mu_r >10\ \text{GeV}$. The stability of the higher orders NLO (dotted-dashed curve), $\text{N}^2$LO (dashed curve) and $\text{N}^3$LO (solid curve) is due to the consideration of coefficients up to $r_{4,j}$ in the determination of the single PMC scale.

Considering for $\Gamma_W^{\text{(had)}}$  the tree level and QCD correction only, i.e. neglecting $\Gamma_\text{EW}$
and $\Gamma_\text{mix}$ [see \eqref{Wdecay}], 
and using $G_F=1.166\times 10^{-5}$ GeV$^{-2}$ \cite{Workman:2022ynf} and $\sum_{i, j}\left|V_{i j}\right|^ 2=1.984 \pm 0.021$ \cite{CMS:2022mhs}, we obtain 
\begin{equation} \label{gammaQCDPMC}
\left. \Gamma^{\text{(had)}}_W  \right\vert_{\text{PMC}}^{\mathcal{O}\paren{\alpha_s^4}}=1405.61\pm 0.21\ \text{MeV}.
\end{equation}

In obtaining this result, the PMC single scale to order next-to-next-leading-Log ($\text{N}^2$LL) is found to be  $Q_s \sim 0.688 M_W$, 
which is smaller than the usual choice of the conventional scale 
$\mu_r \sim M_W$ that is claimed to reduce the large logarithms.

The relative sizes of the LO, NLO, N$^2$L0 and N$^3$LO contributions to $\Gamma_W^{(\text{had})}$ in the PMC method are: 
$\Gamma_{\text{QCD}}^{( \text{LO})}/\Gamma^{(\text{had})}_W \simeq0.039$, $\Gamma_{\text{QCD}}^{(\text{NLO})}/\Gamma^{(\text{had})}_W \simeq 10^{-4}$, 
$\Gamma_{\text{QCD}}^{(\text{N}^2\text{ LO})}/\Gamma^{(\text{had})}_W \simeq -1.5 \times 10^{-3}$ and 
$\Gamma_{\text{QCD}}^{(\text{N}^3\text{LO})}/\Gamma^{(\text{had})}_W \simeq 2.2\times 10^{-5} $.
The uncertainty in Eq.~\eqref{gammaQCDPMC} is obtained by considering the determination of the single scale at NLL and
$\text{N}^2$LL orders. Table \ref{tab1} shows the corrections obtained for each order and the total value at N$^4$LO in the cases of the PMCs method, conventional method and the results obtained in Refs.\, \cite{Kara:2013dua,dEnterria:2020cpv}.


\begin{figure}[]
\centering
\includegraphics[scale=0.65]{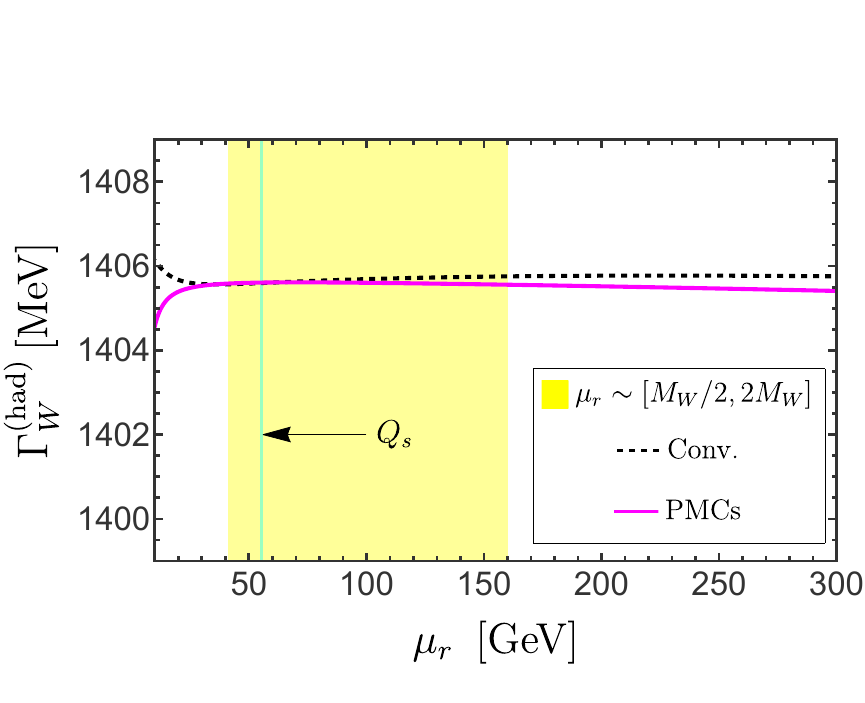}
\caption{Comparison of the results for $\Gamma_W^{\text{(had)}}$ (neglecting $\Gamma_\text{EW}$ and $\Gamma_\text{mix}$), determined to $\mathcal{O}\left(\alpha_s^4\right)$, for the conventional (Conv.) and PMC single scale setting. The highlighted region represents the interval of the renormalization scaling conventions, while the vertical line is the value of single-scale PMC. The input parameters can be found in Section III.}
\label{fig:gammacomparation}
\end{figure}

\begin{table}[]
\centering
\begin{tabular}{lllllll}
\toprule[0.2mm]
\hline
\vspace{0.1cm}  &   $\Gamma_0$  & $\Gamma^{(1)}_{\text{QCD}}$ & $\Gamma^{(2)}_{\text{QCD}}$    & $\Gamma^{(3)}_{\text{QCD}}$ &    $\Gamma^{(4)}_{\text{QCD}}$      & $\Gamma_W^{(\text{had})}$ \\ \hline
PMCs  &          $1352.33$   & $54.898$    & $0.185737$   &                          $-2.10097$ &   $0.302928$  &   $1405.61\pm 0.21$ \\
Conv.   &              $1352.33$   & $51.737$ &     $2.78935$   &                          $-0.966785$ &   $-0.231709$  &   $1405.66_{-5.16}^{+6.36}$ \\
\cite{Kara:2013dua}    &  $1408.98$  &      $54.087$&   $2.727$  &   $-1.018$   &  $-0.245$   & $1464.73$ \\
\cite{dEnterria:2020cpv}    &             $1392.17$ &  $52.345$  &    $2.773$   &                          $-0.925$ &   $-0.221$   &  $1446.15$  \\ \hline
\bottomrule[0.2mm]
\end{tabular}
\caption{Numerical results for the QCD corrections of the hadronic decay of the $W$ boson. $\Gamma_{0}$ represents the contribution at the Born level, $\Gamma_{\text{QCD}}^{(i)}$ is the correction in order $i$, and $\Gamma^W_{\text{QCD}}$ is the total value of the conversion of W to hadrons with QCD corrections. The results are presented in the case of PMC scale-single setting, conventional scale setting (Conv.), and the results presented in other works. The numerical values are given in MeV and the input parameters can be found in Section \ref{sec:numerical}.}
\label{tab1}
\end{table}

Figure \ref{fig:gammacomparation} compares the hadronic decay width of the $W$ boson, $\Gamma_W^{(\text{had})}$, as a function of the renormalization scale $\mu_r$, using both the conventional scale setting and the PMC method. The black dashed curve represents the conventional method, which shows a moderate dependence of $\Gamma_W^{(\text{had})}$ on the renormalization scale within the highlighted region ($\mu_r \in [M_W/2, 2M_W]$). This variation illustrates the inherent sensitivity of the conventional approach to the choice of $\mu_r$, potentially introducing uncertainties in the predicted value. On the other hand, the magenta solid curve corresponding to the PMC single-scale setting displays reduced dependence on $\mu_r$, particularly beyond the region where the scale-setting ambiguity is more pronounced. The vertical line indicates the single-scale value $Q_s$ determined by the PMC, which provides an optimal balance for improving the stability of the perturbative series. While the conventional method can still provide reasonable predictions, the PMC method offers an advantage in terms of reduced scale sensitivity, resulting in a more controlled uncertainty for $\Gamma_W^{(\text{had})}$.

\begin{figure}
    \centering
    \includegraphics[scale=0.65]{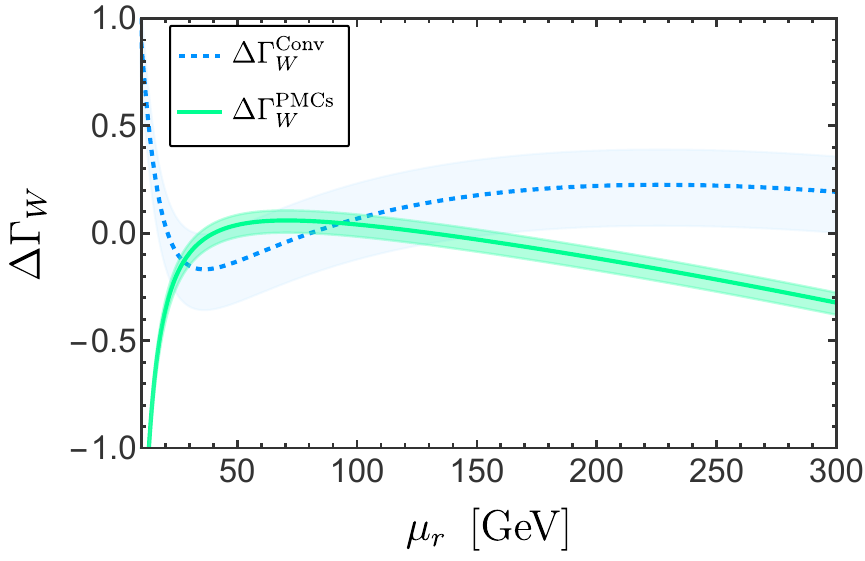}
    \caption{Comparison of the percentage relative variation of the hadronic decay width of the $W$ boson, $\Delta \Gamma_W (\%)$, as a function of the renormalization scale $\mu_r$. The blue dashed curve corresponds to the conventional scale setting, while the green solid curve represents the PMC method. The shaded regions indicate the theoretical uncertainties associated with each method.}
    \label{fig:relative}
\end{figure}

Additionally, to quantify the dependence of the hadronic decay width of the $W$ boson on the renormalization scale, we define the percentage relative variation of the decay width, $\Delta \Gamma_W$, as follows:
\begin{equation}\label{eq:variation}
\Delta \Gamma_W  = \left( \frac{\Gamma_W(\mu_r) - \Gamma_W(M_W)}{\Gamma_W(M_W)} \right) \times 100,
\end{equation}
where this expression represents the relative variation of the decay width $\Gamma_W(\mu_r)$, calculated at an arbitrary renormalization scale $\mu_r$, compared to the reference value $\Gamma_W(M_W)$, evaluated at the scale corresponding to the $W$ boson mass, $M_W$. Eq.~\eqref{eq:variation} quantifies how the renormalization scale affects the theoretical predictions for the hadronic decay width, providing insights into the sensitivity of the calculations to the scale choice, which is a crucial aspect of perturbative QCD predictions. 
Figure \ref{fig:relative} presents the percentage relative variation of the hadronic decay width of the $W$ boson, $\Delta \Gamma_W$, as a function of the renormalization scale $\mu_r$. Two curves are presented to compare different scale setting methods: the blue dashed curve ($\Delta \Gamma_W^{\text{Conv}}$) represents the conventional scale setting method, showing larger oscillations, particularly at lower values of $\mu_r$, indicating higher sensitivity to the scale choice. The solid green curve ($\Delta \Gamma_W^{\text{PMCs}}$) corresponds PMC method, displaying significantly smaller variations and reduced sensitivity to the scale across the entire range. The shaded regions around each curve represent the theoretical uncertainty, with the PMC method also showing smaller uncertainties compared to the conventional approach. This demonstrates the improved reliability of the PMC method in reducing the theoretical uncertainties associated with arbitrary scale choices.

\section{\label{sec:blm}Other optimizations of the renormalization scale}

As we know, a physical quantity can be theoretically estimated as a truncated power series in the coupling $a_s$, which in turn depends on the renormalization scale $\mu_r$ at any given order.  We understand by \emph{optimization of the renormalization scale} $\mu_r$ a method that makes the theoretical expression of a physical quantity as independent as possible on the value of $\mu_r$, as shown above.

The \texttt{Brodsky-Lepage-Mackenzie} 
(BLM) method\footnote{The BLM method is not the only approach to optimize the renormalization scale setting; other methods, such as Fastest Apparent Convergence (FAC) \cite{Grunberg:1980ja} and the Principle of Minimum Sensitivity (PMS) \cite{Stevenson:1980du, Stevenson:1981vj}, have been proposed in the literature with different objectives.}, proposed in 1983
\cite{Brodsky:1982gc}, offers one solution to reduce the ambiguities in the choice of renormalization scale and scheme in pQCD. This approach is inspired by the Gell-Mann-Low scheme used in Quantum Electrodynamics (QED), where the renormalization scale is naturally fixed 
as the momentum of the exchanged photons. 
In the context of QCD, the BLM method provides an automatic way to fix the appropriate renormalization scale, optimizing the convergence of perturbative expansions in pQCD. This method identifies and absorbs terms that depend on the $\beta_0$ coefficient of the beta function into a new or effective running coupling, allowing the perturbative series to be reorganized in such a way that the non-conformal contributions (those that depend on the coefficients $\beta_i$) disappear at the adjusted renormalization scale.
Applying the BLM method to the QCD corrections of the $W$ boson decay $\Gamma_{\text{QCD}}$  in Eq.~\eqref{eq:rNS} up to order $n=2$, the resulting renormalization scale is $Q_{\text{BLM}} = 0.708 M_W$.

For NNLO and higher orders, i.e. $n\geq 3$ in Eq.~\eqref{eq:rNS}, the relationship between $\beta_0$ and $N_f$ is no longer unique. We must extend the BLM method in such a way that we can relate the coefficients that contain powers of 
$N_f$ with the coefficients that contain $\beta_0,\ \beta_1,\ \beta_2$.

A first approach is BLM/PMC \cite{Brodsky:2011ta,Brodsky:2012rj}, which assigns a renormalization scale for each order in the perturbative series pQCD. All non-conformal terms in the perturbative series are summed into the running coupling such that the remaining terms in the perturbative series are identical to those of a conformal theory, namely
$\beta_i =0$ for all $i$.
As a result, the predictions obtained by the BLM/PMC method are independent of the renormalization scheme. The procedure consists of replacing the renormalization scale $\mu_r$ with other effective scales until all higher-order terms with $N_f$ dependence are fully absorbed into the running coupling, thus eliminating the $N_f$ terms associated with the beta function coefficients $\beta_i$. Simultaneously, the perturbative coefficients are modified. In our case, by applying the BLM/PMC method to the perturbative QCD corrections to the hadronic decay of the $W$ boson from Eq.~\eqref{eq:rNS}, we can fix the different effective scales order by order. It should be noted that we lack information about the perturbative values at order $n=5$, so the fourth renormalization scale is fixed to the same value as the last calculated scale. In this manner, $Q_{\text{BLM/PMC}}^{*}=0.740 M_W$ is obtained.

Other approaches include the $\mathcal{R}_\delta$-scheme or PMC method \cite{mojaza2013systematic, brodsky2014systematic} and its particular case of PMCs \cite{singlescale1}, which was used in Section \ref{sec:W-decay}, and the seBLM approach \cite{Mikhailov:2004iq, Kataev:2010du}, all of them as extensions to higher orders of the successful BLM method \cite{Brodsky:1982gc}, which is a method to absorb the $N_f$ terms into an observable at NLO. It is important to note that the first BLM extension was developed by \cite{Grunberg:1991ac}, addressing criticisms regarding how BLM resolves ambiguities. In Ref. \cite{Kataev:2023xqr} an interesting discussion is presented to address the PMC method.

\begin{figure}[]
\centering
\includegraphics[scale=0.65]{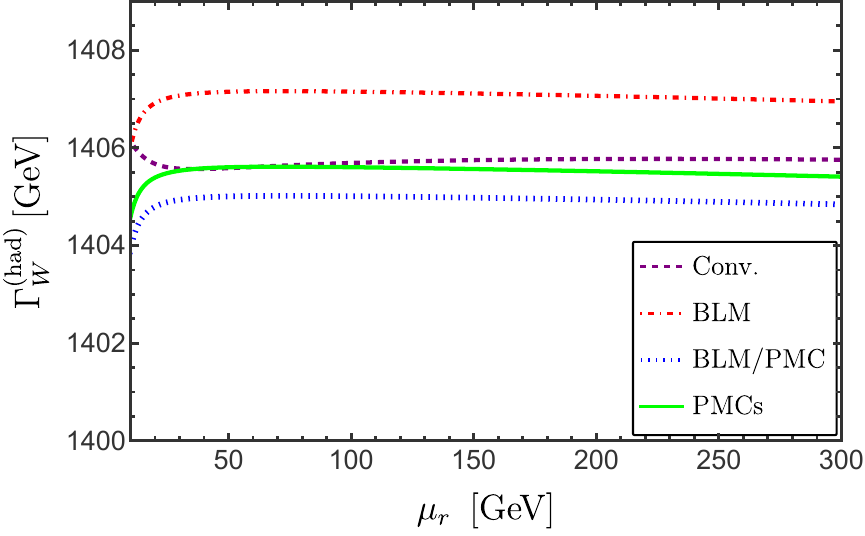}
\caption{Inclusive hadronic decay of $W$ boson $\Gamma_{\text{had}}^{W}$ up a to four-loop level, under conventional scale setting (dash curve), BLM method (dotdashed curve), BLM/PMC method (dotted curve) and PMCs mehod (solid curve). The input parameters can be found in Section \ref{sec:numerical}.}
\label{img3}
\end{figure}

Finally, Figure \ref{img3} shows the estimation of the hadronic decay of the $W$ boson, up to four-loop level, as a function of the renormalization scale $\mu_r$, comparing different scale-setting methods. The dot-dashed curve corresponds to the BLM method, which predicts a larger decay width, while the dotted curve corresponds to the BLM/PMC method, which is slightly lower than BLM but with reduced scale dependence. The dashed curve represents the conventional scale setting, which only stabilizes at higher orders. The solid curve represents the PMC method, which shows the least variation, providing a stable and almost constant prediction over the entire range of $\mu_r$. From this comparison, we can see that the BLM-based methods exhibit less dependence on the renormalization scale, making them a robust choice for accurate QCD predictions.

\section{\label{sec:ckm}Indirect extraction of $\left\vert  V_{cs}\right\vert$}

The hadronic decay width of the $W$ boson, described in Eq.~(\ref{Wdecay}), involves contributions from the elements of the first two rows of the CKM matrix, corresponding to the quarks lighter than the $W$ boson mass. The current experimental values for the CKM matrix elements are as follows \cite{Workman:2022ynf}:
\begin{equation}\label{ckmPDG}
 \begin{array}{ll}
\left|V_{u d}\right|=0.97367 \pm 0.00032, &
 \left|V_{us}\right|=0.22431 \pm 0.00085, \\
 \left|V_{ub}\right|=(3.82 \pm 0.20) \times 10^{-3}, &
 \left|V_{cd}\right|=0.221 \pm 0.004, \\
 \left|V_{cs}\right|=0.975 \pm 0.006,  &
  \left|V_{cb}\right|=(41.1 \pm 1.2) \times 10^{-3}.
\end{array}
 \end{equation}
Of these elements, the most relevant in the hadronic decays of the $W$ boson are $\left|V_{ud}\right|$ and $\left|V_{cs}\right|$. However, the larger uncertainty in the experimental value of $\left|V_{cs}\right|$ introduces a significant limitation to the precision of theoretical predictions for the hadronic decay width. This motivates an independent estimation of $\left|V_{cs}\right|$ using theoretical methods.

The element $V_{cs}$ of the CKM matrix is experimentally determined primarily through processes such as semileptonic decays of $D$ mesons, where the decay width and form factors are combined to extract its value; leptonic decays of the $D_s$ meson, which use the decay rate together with decay constants calculated in lattice QCD; and decays of the $W$ boson, such as $W \to c\bar{s}$, measured at colliders like LEP and LHC \cite{Workman:2022ynf}. These approaches provide values consistent with the unitarity structure of the CKM matrix, albeit with significant uncertainties in the case of $V_{cs}$.

In this work, we applied the PMC method to improve the precision of QCD corrections in the theoretical prediction of the hadronic decay width. Using the theoretical expression for the decay width, along with the experimental values of all parameters, we can indirectly extract the charm-strange CKM element $\left|V_{cs}\right|$, obtaining:
\begin{eqnarray}
\left|V_{cs}\right|_{\text{conv.}} &=& 0.966^{+0.004}_{-0.005}, \\
\left|V_{cs}\right|_{\text{PMC}} &=& 0.9667^{+0.0003}_{-0.0002}.
\end{eqnarray}
The uncertainties shown here were estimated by varying the scale $\mu_r$ in the range $\mu_r/2$ to $2\mu_r$ in the calculation to order N$^3$LO.
These two estimates are consistent with each other and in good agreement with the average value reported by the PDG. The small observed difference highlights the utility of the PMC method in reducing theoretical uncertainties associated with QCD corrections.

\section{\label{summary}Summary}

In this work we have presented an improved determination of the hadronic decay width of the $W$ boson, $\Gamma_W^{\text{(had)}}$, up to the four-loop level using PMCs method. The PMC method systematically eliminates ambiguities in the renormalization scheme and scale, which typically arise in pQCD calculations, providing more precise and stable predictions.

Our results show that the PMC method significantly reduces the dependence of $\Gamma_W^{\text{(had)}}$ on the renormalization scale $\mu_r$, compared to conventional scale setting. As shown in Figures \ref{fig1} and \ref{fig:relative}, PMC-based predictions exhibit much smaller variations in the decay width across the range of renormalization scales considered, indicating reduced sensitivity to higher-order corrections. This demonstrates the advantage of the PMC method in improving theoretical precision for high-energy processes such as $W$ boson decays.

In addition, we have compared to the BLM method and its extensions to higher orders. The BLM/PMC method absorbs all non-conformal terms into the running coupling, further enhancing predictive power by eliminating unnecessary systematic errors. Our results highlight the advantages of these advanced renormalization scale-setting methods over traditional approaches, especially when seeking precision in pQCD predictions.

Finally, the numerical analysis indicates that the PMC method provides a reliable prediction for $\Gamma_W^{\text{(had)}}$ and helps reduce theoretical uncertainties in the determination of other related observables, such as the strong coupling constant $\alpha_s$ and the charm-strange mixing parameter $|V_{cs}|$. The indirect extraction of $|V_{cs}|$ presented here is in good agreement with the latest experimental data, underscoring the relevance of accurate theoretical predictions for understanding the Standard Model parameters.

In conclusion, the PMC method offers a robust framework for improving pQCD predictions, addressing longstanding issues in scale setting and delivering results that are stable and consistent with experimental measurements. This work represents a step forward in the precise determination of the hadronic decay width of the $W$ boson and its role in verifying the robustness of the Standard Model and exploring potential new physics.

\vspace{0.3cm}

\acknowledgments{This work was supported by grants PIIC No. 026/2019 DPP-USM, ANID PIA/APOYO AFB230003 and Fondecyt 1210131 (Chile).}

\appendix

\section{\label{ape1}The PMC reduced perturbative coefficients}
In this appendix, we present the reduced coefficients $\hat{r}_{i,j}$ for the perturbative series of the hadronic decay of the $W$ boson, which can be obtained from the two-point correlator related to the Adler function calculated \cite{Appelquist:1973uz,Chetyrkin:1979bj,Dine:1979qh,Gorishnii:1990vf,Chetyrkin:1996ez,Baikov:2008jh,Baikov:2010je,Baikov:2012er}

\begin{eqnarray}
\hat{r}_{1,0} &=&3C_F, \\
\hat{r}_{2,0} &=&\frac{730}{3}-\frac{121C_{A}}{3T}-176\zeta _{3}+\frac{88C_{A}}{3T}\zeta _{3},\\
\hat{r}_{2,1} &=&\frac{11}{T}-\frac{8}{T}\zeta _{3},
\end{eqnarray}

\begin{eqnarray}
\hat{r}_{3,0} &=&\frac{174058}{9}-17648\zeta _{3}+\frac{8800}{3}\zeta _{5}-%
\frac{1}{T}\left( \frac{172634C_{A}}{27}-77C_{A}^{2}-121C_{A}C_{F}\right.  \\
\nonumber &&\left. -\frac{%
46112C_{A}}{9}\zeta _{3}+56C_{A}^{2}\zeta _{3}+88C_{A}C_{F}\zeta _{3}+4400C_{A}\zeta
_{5}\right) +\frac{1}{T^{2}}\left( \frac{36542C_{A}^{2}}{81}\right. \\
\nonumber  &&\left.+\frac{%
9196C_{A}^{2}}{27}\zeta _{3}\right) -\pi ^{2}\left( \frac{484}{3}-\frac{484C_{A}}{9T}+\frac{121C_{A}^{2}}{27T^{2}}\right) ,\\
 \hat{r}_{3,1} &=&\frac{1}{T}\left( \frac{7847}{9}-\frac{55C_{A}}{2}-\frac{%
33C_{F}}{2}-\frac{2096}{3}\zeta _{3}+20C_{A}\zeta _{3}+12C_{F}\zeta _{3}+%
\frac{200}{3}\zeta _{5}\right) ,\\
\hat{r}_{3,2} &=&\frac{302}{9T^{2}}-\frac{76}{3T^{2}}\zeta _{3}-\pi ^{2}%
\frac{1}{3T^{2}}-\frac{1}{T^{2}}\left( \frac{3322C_{A}}{27}%
 -\frac{836C_{A}}{9}\zeta _{3}\right)\\
 & & -\pi ^{2}\left( \frac{22}{3T}-\frac{11C_{A}}{9T^{2}}\right) ,\\
 \hat{r}_{4,0} &=&\frac{144939499}{81}-\frac{45547960}{27}\zeta_{3}+174240\zeta _{3}^{2}+\frac{527560}{9}\zeta _{5}-\frac{117040}{3}\zeta
_{7} \\
\nonumber &&-\frac{1}{T}\left( \frac{143484077C_{A}}{162} -\frac{109858C_{A}^{2}}{9}-\frac{3157C_{A}^{3}}{24}-\frac{172634C_{A}C_{F}}{9}+\frac{%
1661C_{A}^{2}C_{F}}{3}\right.\\
\nonumber && +\frac{847C_{A}C_{F}^{2}}{2}+\frac{45547960}{27}\zeta_{3}-\frac{2148080C_{A}}{3}\zeta _{3} +\frac{29344C_{A}^{2}}{3}\zeta _{3}+\frac{287C_{A}^{3}}{3}\\
\nonumber &&+\frac{46112C_{A}C_{F}}{3}\zeta _{3}-\frac{1208C_{A}^{2}C_{F}}{3}\zeta_{3}-308C_{A}C_{F}^{2}\zeta _{3} +38720C_{A}\zeta _{3}^{2}-\frac{1305700C_{A}}{27}\zeta _{5}\\
\nonumber &&\left. -\frac{2800C_{A}^{2}}{3}\zeta
_{5}-\frac{4400C_{A}C_{F}}{3}\zeta _{5}-\frac{58520C_{A}}{9}\zeta
_{7}\right) +\frac{1}{T^{2}}\left( \frac{126491101C_{A}^{2}}{772}\right.\\
\nonumber &&\left. -\frac{46508C_{A}^{3}}{27} -\frac{73084C_{A}^{2}C_{F}}{27}-\frac{4919255C_{A}^{2}}{54}\zeta
_{3}+\frac{11704C_{A}^{3}}{9}\zeta _{3}+\frac{18392C_{A}^{2}C_{F}}{9}\zeta
_{3}\right.  \\
\nonumber &&\left.+\frac{4840C_{A}^{2}}{3}\zeta _{3}^{2}-\frac{503360C_{A}^{2}}{27}\zeta
_{5}\right) -\frac{1}{T^{3}}\left( \frac{8160361C_{A}^{3}}{1458}-\frac{270193C_{A}^{3}%
}{81}\zeta _{3}\right. \left.-\frac{13310C_{A}^{3}}{9}\zeta _{5}\right)\\
\nonumber &&  +\pi ^{2}\left( 
\frac{99550}{3}+\frac{49786C_{A}}{3T}-\frac{308C_{A}^{2}}{3T}-\frac{484C_{A}C_{F}}{3T}-\frac{71753C_{A}^{2}}{27T^{2}}+\frac{154C_{A}^{3}%
}{9T^{2}}\right.  \\
\nonumber &&+\frac{242C_{A}^{2}C_{F}}{9T^{2}}+\frac{14641C_{A}^{3}}{108T^{3}}%
+21296\zeta _{3}-\frac{10648C_{A}}{T}\zeta _{3} \left. +\frac{5324C_{A}^{2}}{3T^{2}}\zeta _{3}-\frac{2662C_{A}^{3}}{27T^{3}
}\zeta _{3}\right) ,\\
\hat{r}_{4,1} & = & \frac{1}{T} \left(\frac{13044007}{162}-\frac{78470 C_A}{27}+\frac{649   C_A^2}{18}-\frac{15694  C_F}{9}+\frac{77  C_F^2}{2}+\frac{4367 C_A C_F}{36}\right. \\
\nonumber &&-\frac{195280 \zeta _3}{3} +3520 \zeta_3^2 +\frac{20960 
   C_A\zeta _3}{9}-\frac{236}{9}    C_A^2\zeta _3-28  C_F^2\zeta _3+\frac{4192
    C_F\zeta _3}{3}\\
  \nonumber  && \left.-\frac{794}{9}  C_A
   C_F\zeta _3-\frac{118700 \zeta_5}{27}-\frac{2000}{9}  C_A\zeta _5-\frac{400    C_F\zeta _5}{3} - \frac{5320 \zeta_7}{9} \right)\\
 \nonumber && + \frac{1}{T^2} \left( 
 -\frac{11499191 C_A}{486}+\frac{49075 C_A^2}{81}+\frac{1661 C_A C_F}{3}+\frac{447205  C_A\zeta_3 }{27}-\frac{12350}{27}  C_A^2\zeta_3
  \right. \\
\nonumber && \left. -418  C_A  C_F\zeta _3-\frac{880}{3}   C_A\zeta _3^2+\frac{91520   C_A\zeta _5}{27} \right) + \frac{1}{T^3} \left(  \frac{741851 C_A^2}{486 } -\frac{24563  C_A^2\zeta _3}{27} \right.\\
 \nonumber &&   \left. -\frac{1210  C_A^2\zeta _5}{3}\right)  +\pi^2 \left( -\frac{4526}{3 T}
 +\frac{220 C_A}{9 T}+\frac{44 C_F}{3T}+\frac{968 \zeta _3}{T}+\frac{13046 C_A}{27 T^2}\right.  \\
\nonumber  & & \left. 
-\frac{325 C_A^2}{54 T^2} -\frac{968  C_A\zeta_3}{3 T^2} -\frac{11 C_A C_F}{2 T^2}+\frac{242 \zeta _3 C_A^2}{9
   T^3}-\frac{1331 C_A^2}{36 T^3} \right),
\end{eqnarray}

\begin{eqnarray}
  \hat{r}_{4,2} &=& -\frac{1}{T} \left(  \frac{869 C_A}{72}\frac{121 C_F}{12}-\frac{79  C_A\zeta _3}{9}-\frac{22  C_F\zeta_3}{3}\right) + \frac{1}{T^2} \left( +\frac{1045381}{324}
  \right. \\
  \nonumber && \left. -\frac{3775
   C_A}{27}-\frac{755 C_F}{9}+40 \zeta _3^2+\frac{950  C_A\zeta _3}{9}+\frac{190  C_F\zeta _3}{3}-\frac{40655 \zeta_3}{18} \right. \\
 \nonumber  && \left. -\frac{4160 \zeta _5}{9} \right)+\frac{1}{T^3} \left( \frac{2233 \zeta _3 C_A}{9}+110 \zeta _5
   C_A-\frac{67441 C_A}{162}\right)-\pi^2 \left(\frac{593}{9 T^2} \right.\\
  \nonumber  &&  \left. -\frac{25 C_A}{18 T^2}-\frac{5  C_F}{6 T^2}-\frac{44 \zeta _3 }{T^2}-\frac{121 C_A}{12 T^3}+\frac{22  C_A\zeta _3}{3 T^3}\right),\\
 \hat{r}_{4,3} &=& \frac{1}{T^3} \left(\frac{6131}{54 } -\frac{203 \zeta _3}{3 }-30 \zeta _5  \right) -\pi^2 \left( \frac{11 }{4
   T^3} -\frac{2  \zeta _3}{T^3}\right).
\end{eqnarray}
   
The factors proportional to $\pi^2$ come from the analytic continuation in the Euclidean region as noted in \cite{Pennington:1981cw}. The factors $T$, $C_A$, and $C_F$ are the color factors of the gauge group $SU(3)_C$. The $\zeta_i$ are the Riemann zeta function.


\bibliographystyle{utphys}
\bibliography{refW}



\end{document}